\documentclass[preprint,showpacs,preprintnumbers,amsmath,amssymb,superscriptaddress]{revtex4}

\usepackage{graphicx}
\usepackage{dcolumn}
\usepackage{bm}
\usepackage{color}
\usepackage{hyperref}
\usepackage{subfigure}

\begin{document}
\title{Stochastic and equilibrium pictures of the ultracold FFR molecular conversion rate}
\noindent
\author         {Tomotake Yamakoshi}
\affiliation{Department of Engineering Science, University of Electro-Communications, 1-5-1 Chofugaoka, Chofu-shi, Tokyo 182-8585, Japan}
\author         {Shinichi Watanabe}
\affiliation{Department of Engineering Science, University of Electro-Communications, 1-5-1 Chofugaoka, Chofu-shi, Tokyo 182-8585, Japan}
\author         {Chen Zhang}
\affiliation{Department of Physics and JILA, University of Colorado, Boulder, Colorado 80309-0440, USA}
\author         {Chris H. Greene} 
\affiliation{Department of Physics and JILA, University of Colorado, Boulder, Colorado 80309-0440, USA}
\affiliation{Department of Physics, Purdue University, West Lafayette, Indiana 47907, USA}
\date{\today}

\begin{abstract}
The ultracold molecular conversion rate occurring in an adiabatic ramp through a Fano-Feshbach resonance is studied and compared in two statistical models.
One model, the so-called stochastic phase space sampling (SPSS)[E.Hodby {\it et al}., PRL.{\bf 94} 120402(2005)] evaluates the overlap of two atomic distributions in phase space by sampling atomic pairs according to a phase-space criterion.
The other model, the chemical equilibrium theory(ChET)[S.Watabe and T.Nikuni, PRA.{\bf 77} 013616(2008)] considers atomic and molecular distributions in the limit of the chemical and thermal equilibrium.
The present study applies SPSS and ChET to a prototypical system of K+K$\rightarrow$ K$_2$ in all the symmetry combinations, namely Fermi-Fermi, Bose-Bose, and Bose-Fermi cases.
To examine implications of the phase-space criterion for SPSS, the behavior of molecular conversion is analyzed using four distinct geometrical constraints.
Our comparison of the results of SPSS with those of ChET shows that while they appear similar in most situations, the two models give rise to rather dissimilar behaviors when the presence of a Bose-Einstein condensate (BEC) strongly affects the molecule formation.
\end{abstract}

\maketitle

\section{introduction}
\label{sect:Introduction}
 In recent years, the production and application of ultracold molecules have attracted much attention.
Ultracold polar molecules in particular are being investigated eagerly because of their great promise for numerous applications such as the quantum simulator\cite{Micheli}, ultra high precision measurements\cite{Demille}, ultracold chemical reactions\cite{Ni}, {\it etc}.
However, adaptation of standard laser cooling techniques to molecules is not readily achieved.
Instead experimentalists have employed a scheme to convert ultracold atoms into molecules without heating, {\it i.e.} the combination of Fano-Feshbach resonance(FFR) and Stimulated Raman Adiabatic Passage(STIRAP)\cite{Ni2}.
In this scheme, two atoms get combined to form a quasi-bound molecule in a higher electronic state via a FFR.
The molecules formed get subsequently transferred to the ground state by two color laser pulses.
In this situation, the total production rate is limited by the FFR conversion rate, the transfer efficiency of STIRAP being very high.
Developing an understanding of the FFR phase of the ultracold molecule formation process thus remains important.

The first reported experiment\cite{FCM} swept the external magnetic field to move the unbound low temperature atoms through the FFR region.
According to the experiments, the molecular conversion rate saturates at a very slow magnetic sweep, the so-called adiabatic region.
The adiabatic magnetic sweep is used for the ground state molecular production experiments\cite{Ni2} in order to maximize the number of molecules at a given temperature.
This paper focuses on molecular production at the adiabatic region.
The Landau-Zener(LZ) model has been used for analyzing the experimental results\cite{FCM,Hodby,Voigt,Tyler}. 
Detailed discussions of the LZ model and of its extensions are given in Ref.\cite{sasha}. 

At any rate, Hodby {\it et al}.\cite{Hodby}, developed a semi-classical Monte Carlo simulation method, called the stochastic phase space sampling(SPSS) method to estimate the relation between phase-space density and molecular conversion rate by using a phase-space criterion dependent on a single parameter fitted to the experimental data at one temperature.
No explicit reference to the LZ-type transition mechanism is made in this approach.
Nevertheless the SPSS method has been applied to the $^{40}$K-$^{40}$K\cite{Hodby}, $^{85}$Rb-$^{87}$Rb\cite{Papp}, and $^{40}$K-$^{87}$Rb\cite{Zirbel,olsen} cases, and the results show good agreement with the corresponding experimental data down to the lowest temperature realized in the lab.
We note that recently an experiment on $^{40}$K-$^{87}$Rb molecular production has been carried out in a mixed Bose-Fermi system.
The result of the SPSS model, however, does not agree well with the experimental formation rate of molecules in the region of quantum degenerate temperatures\cite{Tyler}.

On the other hand, Williams {\it et al}.\cite{Williams} developed a theory of the FFR conversion rate based on a coupled Boltzmann-equation treatment that includes atom-molecule interactions.
S.~Watabe and T.~Nikuni\cite{Shohei} extended the theory of Williams {\it et al}. and proposed the so-called (chemical) equilibrium theory (ChET for short in this paper).
The ChET has been applied to noninteracting ideal atomic gases trapped by a harmonic potential, resulting in agreement with the $^{85}$Rb-$^{87}$Rb experiment down to the lowest temperature realized in the lab.
And they commented in Ref.\cite{Shohei} on a possible difference between the SPSS model and the ChET, at temperatures below the BEC-critical temperature $T_c$, a vital region that has not been experimentally examined in detail thus far.
Let us note in passing that no explicit reference to the LZ-type transition mechanism is made in ChET either.

As regards the SPSS model, no systematic assessment of its basic assumptions has been made.
There has been no study of the meaning of the phase-space criterion adopted, nor of possible alternative criteria.
Here we propose a few new phase-space criteria for comparison and analyze the temperature dependence of molecular conversion rate for each criterion.
In doing so, we present an overview of the statistical evaluation of the ultracold molecular conversion rate through comparison between the SPSS and the ChET models.
Furthermore, the molecular conversion rate below $T_c$ is estimated since its experimental and theoretical exploration is much desired.
To understand the dependence on thermodynamic distributions of the atoms and on the phase-space criterion adopted, we choose the number-balanced K$_2$ system in a 3-dimensional harmonic oscillator as a prototype and analyze the features of SPSS in phase space.
The results are compared with the ChET results.
Through these comparisons, we clarify the distinct temperature dependence of the molecular conversion rate between the SPSS and ChET models qualitatively.
Number-imbalanced systems are also considered for other effects such as the gravitational sag as well as the gap between two different values of $T_c$.
Incidentally, the ChET model with interacting atoms has been developed in Ref.\cite{Nishimura}. 
This current paper, however, focuses on noninteracting systems only to understand the overall behavior. 

The paper is organized as follows.
 Sec.\ref{sect:method} outlines the SPSS and ChET models and gives some additional background. 
Sec.\ref{sect:K2} discusses the results of the SPSS model for the prototypical K$_2$ case with four different phase-space criteria.
Sec.\ref{sect:other} shows the results for Rb$_2$ and KRb cases, systems for which the temperature dependence of the conversion rate has been investigated experimentally.
Sec.\ref{sect:equ} compares the results of the SPSS model with the ChET.
Sec.\ref{sect:conclusions} concludes the paper.
This paper uses $T_c=$ 0.94 $\hbar \bar{\omega} N^{1/3} / k_B$ for the unit of temperature throughout, where $\hbar$, $\bar{\omega}$, N, and $k_B$ correspond to the
Planck constant, mean trapping frequency, number of bosonic atoms, and the Boltzmann constant respectively.
And we define the ``molecular conversion" rate as the number of formed molecules divided by the initial number of minority atoms, namely $\chi_{m}=N_{\rm molecule}/N_{\rm atom,minority}$.

\section{Theoretical Models: SPSS and ChET}
\label{sect:method} 
The theory of molecular formation in a cold mixture of two atomic species is based on the following effective Hamiltonian as a common starting point, namely\begin{eqnarray*}
{\cal H}&=&\int\ d{\bf r}[ \sum_{\alpha=1,2}\hat\psi_\alpha^\dagger({\bf r}){\cal H}_A^{(\alpha)}\hat\psi_\alpha({\bf r})
+\hat\phi_M^\dagger({\bf r}){\cal H}_M\hat\phi_M({\bf r})]\\
&&+\kappa\int\ d{\bf r}\ \{\hat\phi_M^\dagger({\bf r})\hat\psi_B({\bf r})\hat\psi_A({\bf r})+{\rm h.c.}\}\\
&&+g_A\int\ d{\bf r}\ \hat\psi_A^\dagger({\bf r})\hat\psi_B^\dagger({\bf r})\hat\psi_B({\bf r})\hat\psi_A({\bf r})
\end{eqnarray*}
where $\hat\psi_A$ and $\hat\psi_B$ are the atomic field operators for the two types of atoms, and $\hat\phi_M$ is the molecular field operator\cite{QFT};
\[
{\cal H}_A^{(\alpha)}=-\frac{\hbar^2}{2m_\alpha}\nabla^2+U_\alpha({\bf r})
\]
and
\[
{\cal H}_M=-\frac{\hbar^2}{2(m_A+m_B)}\nabla^2+U_M({\bf r})+\epsilon_{\rm res}
\]
represent the single-particle Hamiltonian for atoms and molecules, respectively. Here the atomic masses $m_A$ and $m_B$ differ in general. $U_\alpha({\bf r})$ and $U_M({\bf r})$ are the external potentials for the trap and gravity, respectively. The energy 
$\epsilon_{\rm res}$ of the resonant molecule is tuned close to a collision threshold in experiments by varying an external magnetic field ${\bf B}$. 
The constants $\kappa$ and $g_A$ pertain to the association/dissociation of an atomic pair into/from a molecule, and the atom-atom interaction, respectively.
The LZ-type transition leads to $\kappa$, but both SPSS and ChET bypass the use of the interaction term by employing physically-motivated assumptions.
The atom-atom interaction term is dropped in what follows.
In the mixture of two types of atomic gas, the interchannel transition for molecular formation is effective when the magnetic field sweeps across the resonance energy. 
The experimentally observed and reported molecular conversion rate at a given temperature pertains to the saturated conversion rate that is measured at a reasonably slow magnetic sweep rate.
Let us consider first the moment when the magnetic field is far away from the Fano-Feshbach resonance region so that the relevant classical Hamiltonian is that of an aggregate of independent atoms in the harmonic oscillator trapping potentials. 
The quantum distribution functions are given by
\begin{equation}
f_\alpha=\frac{1}{\exp[{(H_\alpha-\mu_\alpha)/k_B T}]\pm 1}.
\label{eq:f}
\end{equation}
where the Hamiltonian for a single atom of type $\alpha$ is
\begin{equation}
H_\alpha = \frac{1}{2m_\alpha}\vec{p}^2 + \frac{1}{2}m_\alpha \omega_\alpha^2 \vec{r}^2 ,
\end{equation}
where ``$+$" corresponds to fermions, and ``$-$"  to {\it thermal} bosons.

We extend the previously developed SPSS in order to treat the condensate bosons (BEC) using the truncated Wigner approximation to the phase-space distribution. 
The ChET on the other hand does not require a specific distribution function for BEC but rather, only the chemical potential. 
Let us employ the ground state wave function of the harmonic oscillator to represent a noninteracting BEC and apply the Wigner transformation using
\begin{equation}
\phi_0(\vec{r})\propto\exp \left( {-\frac{m_\alpha\omega_\alpha}{2\hbar}\vec{r}^2} \right).
\end{equation}
This yields
\begin{eqnarray}
P_\alpha(\vec{r},\vec{p})&=&\frac{\hbar^3}{2\pi^3}\int_V \phi_0^\ast \left( \vec{r}+\frac{\vec{q}}{2} \right)\phi_0 \left( \vec{r}-\frac{\vec{q}}{2} \right) e^{i\vec{p}\cdot\vec{q}/\hbar}\ d\vec{q}
\nonumber\\
&\propto&\exp \left( {-\frac{\vec{p}^2}{m_\alpha\omega_\alpha\hbar}-\frac{m_\alpha\omega_\alpha}{\hbar}\vec{r}^2} \right).
\label{eq:BEC-Wig}
\end{eqnarray}
This probability distribution does not depend explicitly on the number of atoms unlike the 
Thomas-Fermi(TF) wave function \cite{Papp}.
On the other hand, the number of BEC atoms in the harmonic oscillator trap is given analytically by
\begin{eqnarray}
N_\alpha^0 &=& \left[1-\left(\frac{ T}{ T_c}\right)^3\right] N_\alpha,
\label{eq:N0}
\end{eqnarray}
where $N_\alpha^0$, and $N_\alpha$ correspond respectively to the number of BEC atoms and the total number of atoms of type $\alpha$ at given temperature $T$. 
The probability of finding a condensate atom is then $N_{\alpha}^{0}P_{\alpha}(\vec{r},\vec{p})$.

Before going farther, note that the BEC transition temperature $T_c$ differs in general for the two types of atoms due to the mass difference.
The TF (Thomas-Fermi) wave function for a repulsively interacting BEC tends to spread out more and get flatter near the center of the trap so that the distribution tends more toward smaller momentum values than in the noninteracting Eq.~(\ref{eq:BEC-Wig}). 
To calculate the molecular conversion rate, Papp and Wieman\cite{Papp} represented the effect of BEC by assuming that if a pair of atoms find themselves within 
the TF radius, they then form a molecule regardless of their relative momentum. 
Thus Ref.~\cite{Papp} does not generate a phase-space distribution. 
Instead, they use only the spatial extent of the TF distribution without setting any criterion on the momentum. 

\subsection{SPSS: Stochastic Phase Space Sampling}

The SPSS is a method that samples candidate atoms for molecular formation by semi-classical Monte Carlo simulation.   Ref.~\cite{Hodby} imposed a certain reasonable-looking phase-space constraint on the initial atom-pair distribution.
The conversion rate is calculated under the following two assumptions.
\begin{enumerate}
\item Only the atomic pairs satisfying a given phase-space criterion can form a molecule.
\item Once a molecule is formed, it does not dissociate into atoms.
Indeed, the lifetime of a molecule is known to be considerably longer than the time scale of an experiment\cite{Regal}.
\end{enumerate}

\begin{figure}[htbp]
 \begin{center}
  \includegraphics[width=80mm]{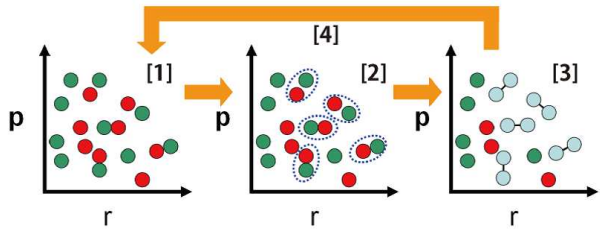}
 \end{center}
 \caption{(Color online) Conceptual diagram representing ``Stochastic Phase Space Sampling (SPSS)''. See steps [1]-[4] described in the text.}
 \label{fig:1}
\end{figure}

A simulation is implemented in the following four steps, illustrated in Fig.~\ref{fig:1}:
\begin{enumerate}
 \item Distribute atoms randomly in phase space in accordance with the thermodynamic equilibrium distribution at the initial temperature $T$, Eq.(\ref{eq:BEC-Wig}). This temperature is fixed, meaning that the process is regarded as isothermal.
 In the presence of a BEC, Eq.(\ref{eq:N0}), is used for the condensed atoms.
 \item Search for atomic pairs that satisfy the phase-space criterion. Avoid double-counting by erasing the used atoms from the list of candidates.
 \item Count the number of atomic pairs found in the random search. Regard it as the number of the formed molecules.
 \item Return to Step [1], and repeat the steps until the statistical noise is reduced below the target noise level.\\
\end{enumerate}

Since the phase-space volume made available by a geometrical constraint plays a key role in SPSS, one of our goals is to assess the sensitivity of the method to the criterion adopted for pair formation.
In what follows we define four independent phase-space criteria, the first one being the version initially employed by
Ref.\cite{Hodby} which yielded an extremely good fit to the experimental results for the Bose gas and for a two-component Fermi gas. 
As we will see in the discussion in Sec.~\ref{sect:discussions}, all these separate criteria reproduce
the temperature dependence of the experimental molecular conversion rate satisfactorily at thermal temperatures, but show some departures below
the BEC critical temperature. 

Let us introduce the following four phase-space criteria;\\
Criterion 1:
\begin{equation}
|\Delta r \Delta p| < \gamma_q h,
\label{crit-1}
\end{equation}
Criterion 2:
\begin{equation}
|\Delta \vec{r} \times \Delta \vec{p}| < \gamma_l \hbar ,
\label{crit-2}
\end{equation}
Criterion 3:
\begin{equation}
(|\Delta x \Delta p_x||\Delta y \Delta p_y||\Delta z \Delta p_z|)^{1/3} < \gamma_v h 
\label{crit-3}
\end{equation}
Criterion 4:
\begin{equation}
|\Delta x_i \Delta p_i| < \gamma_s h \quad(i=x,y,z)
\label{crit-4}
\end{equation}
where $\Delta r$ corresponds to the spatial separation of an atom pair and $\Delta p=m_{{\rm rm}} \Delta v$, where $m_{{\rm rm}}$ equals the reduced mass, 
$\Delta v$ corresponds to the relative velocity, and $\gamma_q$, $\gamma_l$, $\gamma_s$, and $\gamma_v$ are the pairing parameters which are assumed to be independent of temperature and density. 
Criterion 1 is introduced on the basis of the number of accessible quantum states in the relative coordinate. 
Criterion 2 is introduced under the premise that the low partial waves are important for molecular formation.
The use of $\hbar$ instead of $h$ for Criterion 2 is thus intentional. 
Criterion 3 is similar to criterion 1, but represents the full phase-space volume occupied by the pair.
Criterion 4 is introduced as a counterpart to criterion 3, actually a most familiar form in statistical mechanics. 
In any case, there is no {\it a priori} numerical or theoretical support for any one of these criteria in the present context of molecular formation. 

As for the accuracy targeted in the present paper, the simulation program generates probability distributions, as given above by Eq.(\ref{eq:f}), by the  rejection sampling, 
and then
searches for all pairs that meet the phase-space criterion. 
The period of the random number generator is about $2.3\times10^{18}$ for this study. This is sufficient for dealing with $10^{6}$ atoms.
The search ends when no remaining atomic pair satisfies the criterion.
These searches are repeated a sufficient number of times, and the results are averaged. 
The error is of the order of the inverse of the square root of the number of atoms times the number of iterations. 
Throughout our calculations this error is reduced to within 1 \%.
We shall return to the discussion of these criteria in Sec.~\ref{sect:discussions}.

\subsection{ChET: Equilibrium Theory}

The ChET was developed by E.Williams {\it et al}.\cite{Williams}, and extended further by S. Watabe {\it et al}.\cite{Shohei}.
It is based on the result of the coupled atom-molecule Boltzmann equation approach.
In the ChET, one solves simultaneous equations consistently with equilibrium conditions to obtain the molecular conversion rate.
The first equation is the conservation of the total number of atoms,
\begin{equation}
N_{tot}=N_A+N_B+2N_M
\end{equation}
and the second equation states the constancy of the number of each atomic species, hence
\begin{equation}
\alpha=\frac{N_B+N_M}{N_A+N_M}
\end{equation}
where $N_A$, $N_B$, and $N_M$ denotes the number of the majority atoms, of the minority atoms, and of the molecules, respectively.
And the ratio $\alpha$ is defined by the initial ratio $N_B/N_A$, where we consider $N_A \geqq N_B$ , where $N_M=0$ thus $1 \geqq \alpha$. 
There are two equilibrium conditions, one representing chemical equilibrium,
\begin{equation}
\mu_A + \mu_B = \mu_M + \delta
\label{eq:CPE}
\end{equation}
where $\mu$ denotes the chemical potential of each component, and $\delta$ denotes the detuning, namely the energy difference between dissociated atomic state and the molecular bound state.
The other condition represents thermal equilibrium, $T_A=T_B=T_M$, where $T$ denotes the temperature of each component.
The population of each component is a function of $\mu$ and $T$.
ChET uses two assumptions.
One is that the magnetic sweep process is adiabatic, thus the total entropy of the system is conserved.
And the other one is that molecular production halts at $\delta=$0, due to the conservation of momentum and energy.
So the method first calculates the total entropy at $\delta \rightarrow \infty$ at a given $T$ and $\alpha$, and then traces the adiabatic state until $\delta=$0.
Thus the molecular conversion rate is given by equating the total entropy at $\delta \rightarrow \infty$ and that at $\delta=$0,
\begin{equation}
S_A(T_{\infty},\mu_{\infty,A})+S_B(T_{\infty},\mu_{\infty,B})=S_A(T_0,\mu_{0,A})+S_B(T_0,\mu_{0,B})+S_M(T_0,\mu_{0,A}+\mu_{0,B}),
\label{eq:entro}
\end{equation}
where S denotes the entropy of each component. 

\section{Discussions}
\label{sect:discussions}

 Features of the molecular conversion rate are now explored for three statistically distinct types of atom-atom pair, namely Fermi-Fermi, Bose-Bose, and Bose-Fermi.
 To this end, we employ K$_2$ as a prototype for our analysis, considering $^{39}$K for the boson, and $^{40}$K for the fermion. 
The K$_2$ system simplifies the situation on two accounts. 
Firstly, there is negligible gravitational sag thanks to the nearly equal masses of the isotopes. 
Secondly, we may presume that there are no more than two characteristic temperatures in the K$_2$ system, namely the Fermi temperature $T_F$ and the Bose-Einstein condensation point $T_c$\cite{K2C}.
It is also worthwhile pointing out that in a theoretical proposal as well as in a recent experiment with $^{40}$K-$^{87}$Rb\cite{Silke}, the authors of the experimental study employed an optical dipole trap with the so-called ``magic frequency'', which eliminates the gravitational sag. 
For this reason, the molecules $^{39}$K-$^{40}$K and $^{40}$K-$^{87}$Rb can be considered equivalent under an appropriate scaling of parameters\cite{Silke}. At any rate, the issues gravitational sag and the gap in $T_c$ are explored as supplementary items later in Subsection B.

Let us now compare the results of SPSS models and that of the ChET. To begin with, observe
that the ChET is based on entropy conservation during an adiabatic magnetic sweep and on the condition that 
the statistical distribution of atomic pairs and that of molecules are always in equilibrium. 
The SPSS model presumes the conservation of the two-body local phase-space volume but no equilibrium.
However, when atomic pairs are chosen the SPSS appears to be adiabatic since conservation of the phase-space volume amounts to adiabatic invariance and thus to conservation of entropy.
But consistency with thermal equilibrium is not met.
In the presence of an atomic BEC in the Bose-Bose and Bose-Fermi cases, ChET assumes the chemical potential of the bosonic atom component is 0, so that the conversion rate gets flat-lined. 
However, in the presence of two atomic BECs in the Bose-Bose case, ChET reveals certain temperature dependence of the conversion rate.
For the sake of comparison, we determine the molecular conversion rate using the procedure developed in Ref.~\cite{Shohei} and review it in Sec.~\ref{sect:equ}.

\subsection{Features with Prototypical K$_2$ Molecules}
\label{sect:K2}

The following specific systems are considered.
\begin{itemize}
\item  Fermi-Fermi: $^{40}$K($f=\frac{9}{2},m_f=-\frac{9}{2}$)-$^{40}$K($f=\frac{9}{2},m_f=-\frac{7}{2}$)
\item  Bose-Bose: $^{39}$K-$^{39}$K
\item  Bose-Fermi: $^{39}$K-$^{40}$K
\end{itemize}
Note that experimental data on K$_2$ formation is known to us only for the Fermi-Fermi system\cite{Hodby}. 

\begin{table}[h]
\caption{Experimental parameters from \cite{Hodby} used in the present simulation. $N_{\rm K}$ is the number of K atoms, $\omega_{\rm trap}$ the trap frequency, and $\gamma$'s pertain to the criteria.}
\begin{tabular}{cccccc}
\hline\hline
$N_{\rm K}$  &$\omega_{\rm trap}$ & $\gamma_Q$ &$\gamma_L$ &$\gamma_V$ &$\gamma_S$\\
\hline
3$\times$ 10$^{4}\:$  & $\:$ 2$\pi\times $(470, 470, 6.7) Hz $\:$ & 0.19 $\:$ & $\:$ 0.26 $\:$ & $\:$ 0.0085 $\:$ & $\:$ 0.050\\
\hline\hline
\end{tabular}
\label{tab:parameters}
\end{table}

For the isotropic trap considered in this paper, we rescale the length and momentum according to the
trap frequencies. 
The values of $\gamma_q$, $\gamma_l$, $\gamma_v$, and $\gamma_s$ are thus determined by the fitting to the experimentally available 
$^{40}$K($f=\frac{9}{2},m_f=-\frac{9}{2}$)$^{40}$K($f=\frac{9}{2},m_f=-\frac{7}{2}$) data, and are used for the other Bose-Bose and Bose-Fermi cases as well.
Table~\ref{tab:parameters} shows the values of various parameters employed in this section for K$_2$. 
The accessible phase-space volume for the atomic pairs increases as $\gamma$ increases.
The fitting parameters thus become smaller in the order of $\gamma_Q > \gamma_S > \gamma_L/2 \pi > \gamma_V $.

\subsubsection{Temperature dependence of the conversion rate}
\label{sect:TDC}
\begin{figure*}[ht]
  \includegraphics[width=12cm]{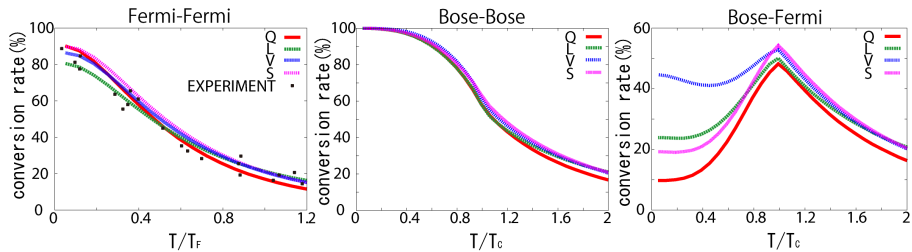}
  \caption{(Color online) SPSS molecular conversion rates for four criteria shown altogether. 
  The red line corresponds to the result of Criterion 1, green to Criterion 2, blue to Criterion 3, and pink to Criterion 4. 
  From left to right the panels correspond to $^{40}$K-$^{40}$K, $^{39}$K-$^{39}$K, and $^{40}$K-$^{39}$K, respectively.}
  \label{fig:all3}
 \end{figure*}

Stereotypical behavior of the conversion rate is summarily displayed as a function of temperature in Fig~\ref{fig:all3} for the four criteria altogether.
The thermal limits look similar for the three types of systems. 
It is probable that this is due to the fact that the Hamiltonian and the geometrical constraints are 
all quadratic in $r$ and $p$. And also the chemical potential plays little role in the thermal limits. 
Then the relevant quantities can all be expressed in terms of 
temperature-scaled distance and momentum, namely $\tilde r=r/\sqrt{T}$ and $\tilde p=p/\sqrt{T}$. The magnitude of
the formation rate depends then on the value of $\gamma$/$T$ which controls the phase-space volume. 
Specific temperature dependences become apparent when the chemical potential $\mu$($T$) plays a role, 
namely at $T\leq T_c$ for boson and at $T\leq$0.6$T_F$ for fermion where the chemical potential changes sign. 
See Fig.\ref{fig:all3}.
In the Bose-Bose case the conversion rate grows close to 100\% as $T\rightarrow 0$, suggesting complete overlap of the two BECs in this limit which makes it easy to satisfy the imposed criterion.
However, the slope of the molecular conversion rate is discontinuous at $T_c$ because of the BEC phase transition across this temperature. 
In the Fermi-Fermi case, the conversion rate rises monotonically as $T$ lowers.
The $T$=0 limit is finite, however.
No discontinuity in the slope occurs at $T_F$ since there is no phase transition. (Note that the Fermi temperature $T_F$=1.93  $T_c$ is out of the range displayed in this figure.)
The Bose-Fermi case shows a noticeable discontinuity in slope because of the BEC phase transition and of the subsequent diminishing overlap in phase space. 
 
\subsubsection{Relevant single- and two-particle  phase-space regions}
\label{sect:RPS}
 Let us begin this subsection with the following question: 
Should the four alternative criteria pick out similar regions of 
the phase space when $\gamma_Q$, $\gamma_L$, $\gamma_V$, 
and $\gamma_S$ are each made to fit to the observed molecular conversion rate? 
Let us look into the initial phase-space distribution of atoms that lead to molecular formation 
by comparing the results of the four criteria.
 \begin{figure}[h]
  \includegraphics[width=8cm]{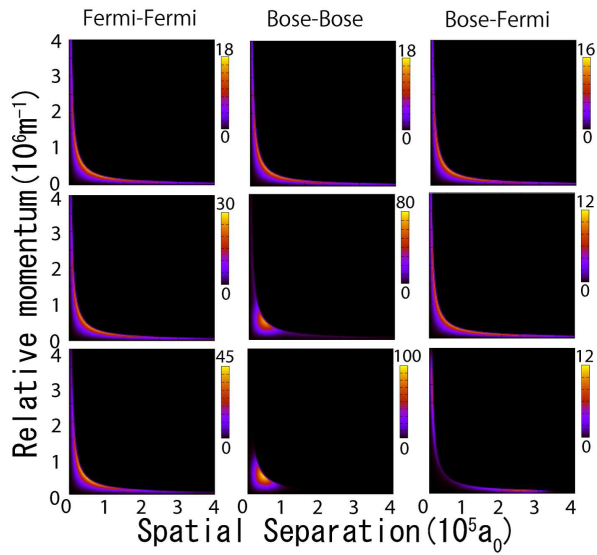}\\  
  
\caption{(Color online) Master diagram for Q(Criterion1) showing phase-space distribution of molecule-forming pairs, $\rho(\Delta r,\Delta p)$ Eq.~\ref{eq:integ}.  
Each consists of three panels corresponding to $ T> T_c$, $ T\sim T_c$, and $ T< T_c$. 
The triplet of distributions are arranged so that the left one is the Bose-Bose case, the middle one the Fermi-Fermi case, and the right one the Bose-Fermi case. 
The horizontal axis shows the relative spatial separation $\Delta r $, while the vertical axis shows the relative momentum $m_{red} \times \Delta v$. The brighter the color, the higher the density.  
$\gamma_q=$0.19 as in Table.~\ref{tab:parameters}.}
  \label{fig:Q_Master}
 \end{figure}

 \begin{figure}[h]
    \includegraphics[width=8cm]{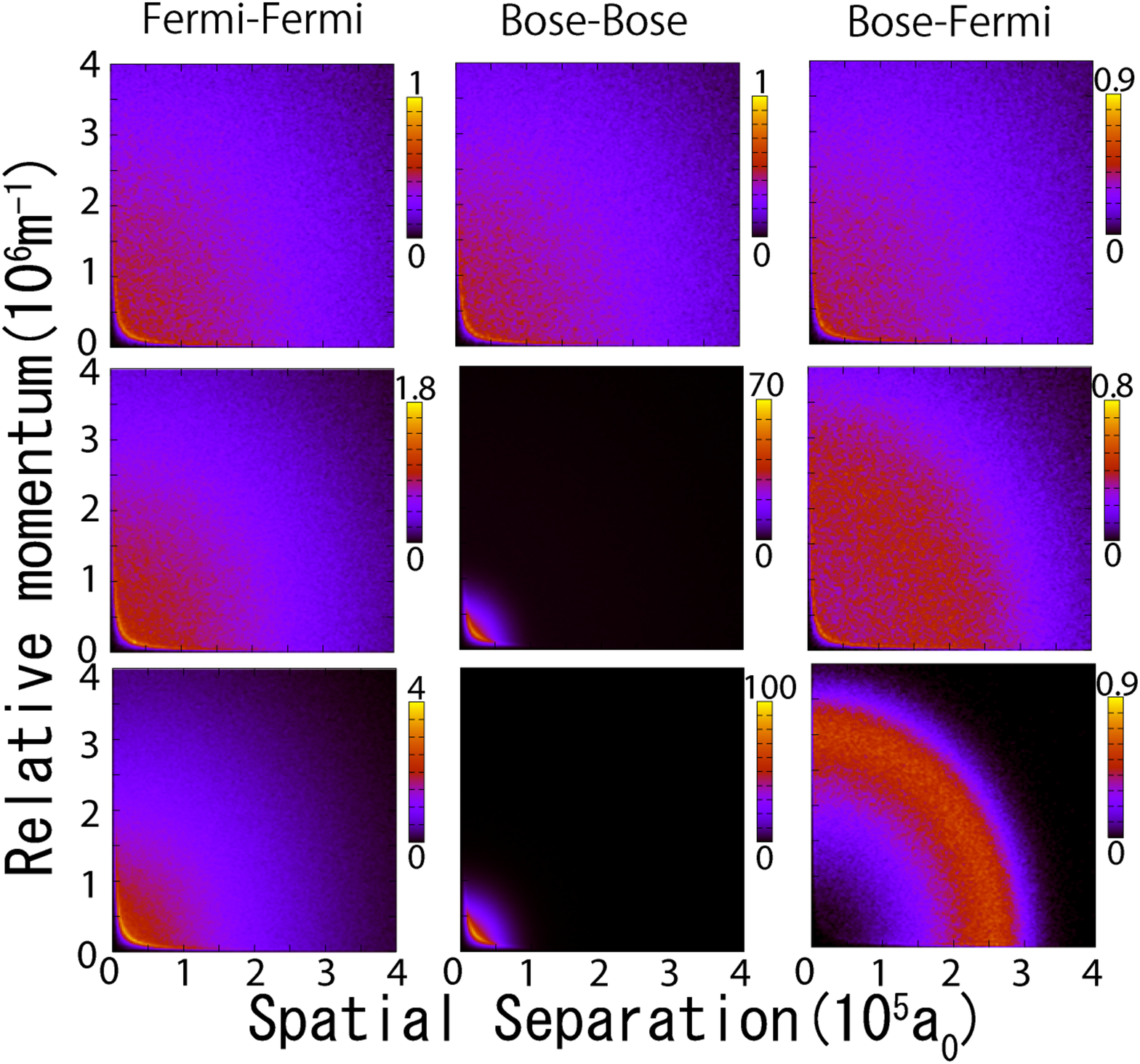}\\ 
  \caption{(Color online) Master diagram for L(Criterion2) showing phase-space distribution of molecule-forming pairs, $\rho(\Delta r,\Delta p)$ same as Fig.~\ref{fig:Q_Master}.
  $\gamma_q=$0.26 as in Table.~\ref{tab:parameters}.}
  \label{fig:L_Master}
 \end{figure}
 
 \begin{figure}[h]
  \includegraphics[width=8cm]{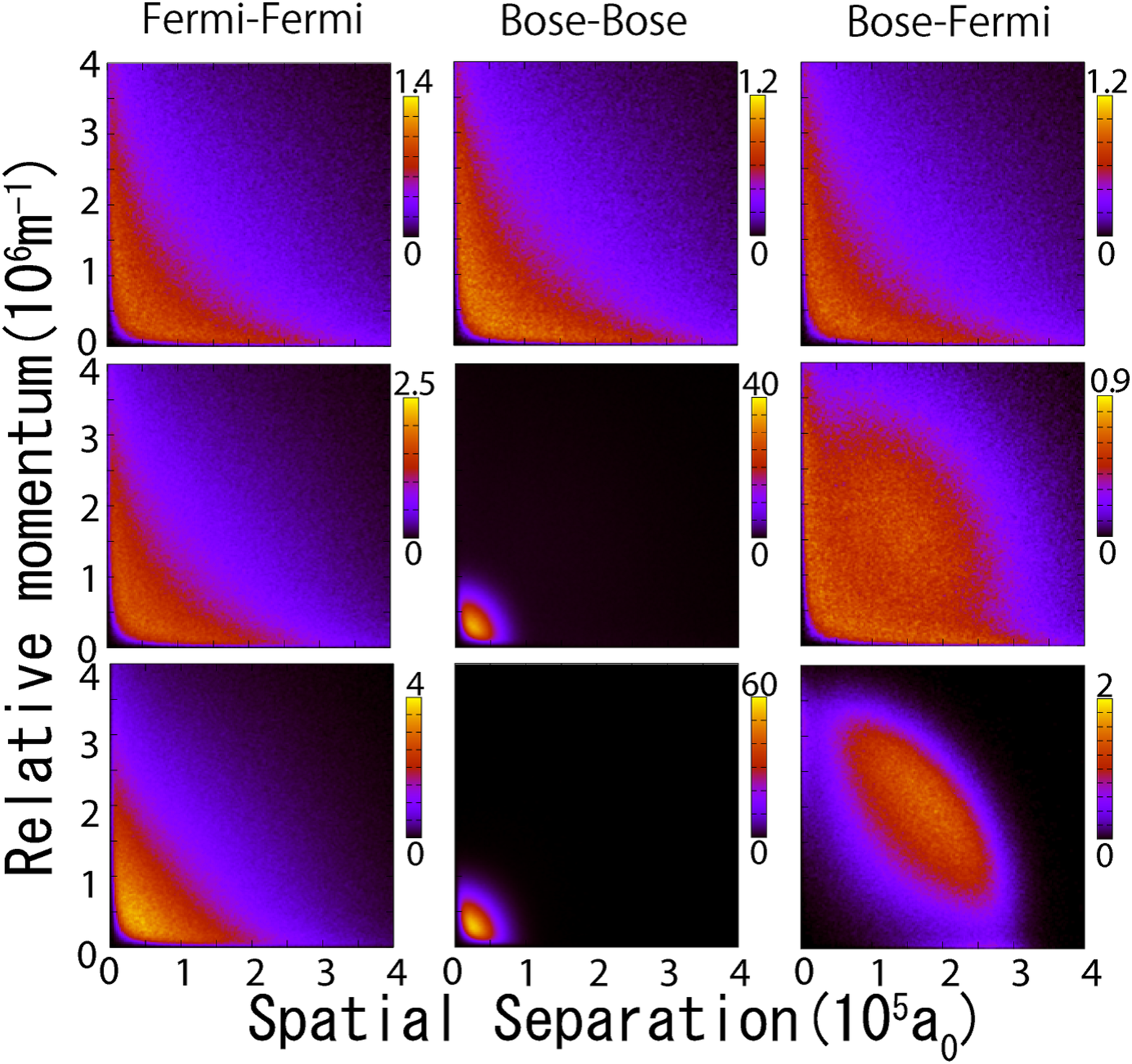}\\ 
  \caption{(Color online) Master diagram for V(Criterion3) showing phase-space distribution of molecule-forming pairs, $\rho(\Delta r,\Delta p)$ same as Fig.~\ref{fig:Q_Master}. 
$\gamma_q=$0.0085 as in Table.~\ref{tab:parameters}.}
  \label{fig:V_Master}
 \end{figure}

 \begin{figure}[h]
  \includegraphics[width=8cm]{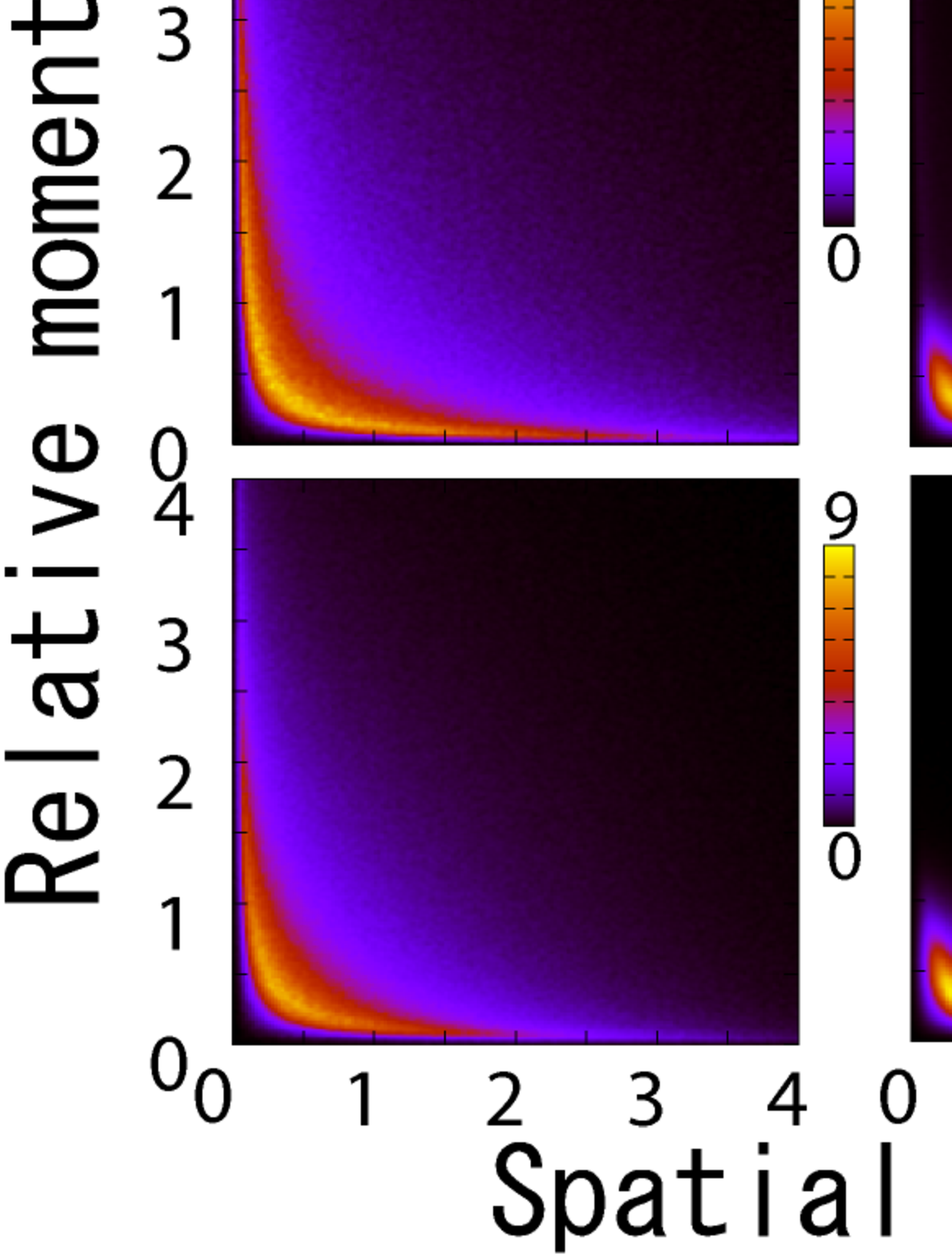}\\ 
  \caption{(Color online) Master diagram for S(Criterion4) showing phase-space distribution of molecule-forming pairs, $\rho(\Delta r,\Delta p)$ same as Fig.~\ref{fig:Q_Master}.
$\gamma_q=$0.050 as in Table.~\ref{tab:parameters}.}
  \label{fig:S_Master}
 \end{figure}

To this end, we define the ``two-particle" distribution function represented by the number of candidate pairs
 averaged over the implemented iterations, $N_{\rm iter}$,
within a suitably chosen phase-space domain $D(\Delta r,\Delta p)$ of 
a moderately small volume in the neighborhood of $(\Delta r,\Delta p)$,
\begin{equation}
\rho(\Delta r,\Delta p)=\frac{1}{N_{\rm iter}}\sum_j N[(\Delta r_j,\Delta p_j)\in D(\Delta r,\Delta p)]
\label{eq:integ}
\end{equation}
Figs.~\ref{fig:Q_Master}, \ref{fig:L_Master}, \ref{fig:V_Master}, and \ref{fig:S_Master} are 
our master diagrams showing $\rho(\Delta r,\Delta p)$ comprehensively. 
Each panel consists of the three symmetry cases. 
Down each column, the temperature varies from $ T> T_c$ to $ T< T_c$. 
And $a_0$ represents the Bohr radius.

 Next consider the physical implications of the way the candidate pairs are distributed. 
First, we consider how the candidate pair distributions differ, depending on the combination of atomic species.
In the Fermi-Fermi case, the shape of the candidate pair distributions remains unchanged throughout 
these temperature regions.
In the Bose-Bose case, we can see a change in the pair distribution below $T_c$, namely a reduction in size toward the origin. 
This change comes from the condensation of the bose gases where their phase space volumes become very small.
In the Bose-Fermi case, there are special features near zero temperature. 
And these features depend critically on the phase-space criterion. 
Above $T_c$ on the other hand, there are no marked differences in the candidate pair distributions for the three symmetry cases because they are all representable by the Maxwell-Boltzmann distribution.
One major difference throughout the diagram is that Criteria 2, 3, and 4 permit more widely spread distributions than does Criterion 1. 
Stated somewhat differently, the distributions simulated by Criteria 2, 3, and 4 are generally sparse 
with their maxima roughly in the regions cut away sharply by Criterion 1. 
The sharp cut-aways with respect to Criteria 2, 3, and 4 do not introduce sharp edges in this $\Delta r$-$\Delta p$ representation unlike Criterion 1. 
Incidentally, the converse is not necessarily true. Let us note that in the phase space of angular momentum and relative kinetic energy, both Criterion 1 and 2 show sharp edges. 
The maximum value of the angular momentum corresponding to the edge is about 1.2$\hbar$ so that the s-wave contribution is dominant, but with nonnegligible p-wave contribution.
In all the panels, the high concentration of most likely candidates appear along similar-looking hyperbolic-shaped arcs near the origin for the Bose-Bose and Fermi-Fermi cases if on different scales. 

 \begin{figure}[h]
 \includegraphics[width=8cm]{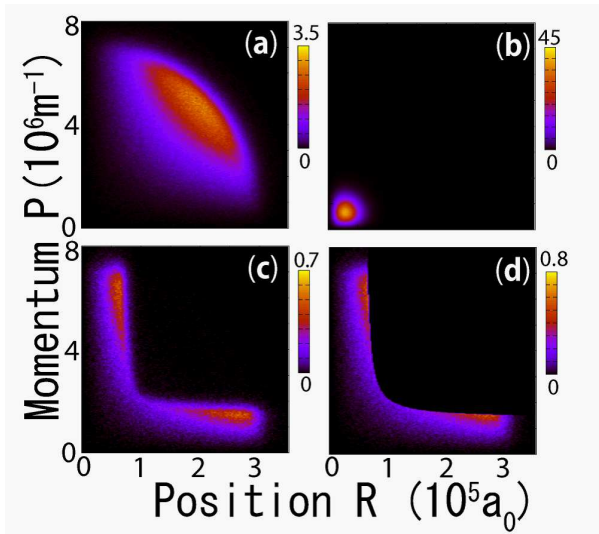}
  \caption{(Color online) The upper two figures represent phase-space density distributions of individual components; (a) the fermion to the left and (b) the boson to the right at $T$=0.1$T_c$.  
  (c) shows the density of the molecule-forming pairs under Criterion 1. 
  (d) shows the density of the fermion cut away by Criterion 1 when the boson is assumed to be concentrated entirely at the origin.
  These two resemble each other.}
  \label{fig:ind-1}
 \end{figure}

On the other hand, the Bose-Fermi case exhibits a considerably different appearance.
To aid in understanding, plotted in the two upper panels of Fig.~\ref{fig:ind-1} are the phase-space 
density distributions of the individual atoms in single-particle phase space. The one on the left represents the fermion, and the
right one the boson which is condensed near the origin. As a result, the molecular formation is restricted to the
narrow overlapping regions of these single-particle distributions.
The lower two panels of the same figure mark
the fermionic atoms that contribute to the molecular formation simulated by Criterion 1. The one on the left shows the distribution of 
the fermion atoms in the formed molecules which is extracted in the course of numerical simulations. The one on the right
shows the same distribution but constructed by applying Criterion 1 to the single-particle distributions of the upper two panels.
We note that they look rather similar.
The boson (BEC) distribution being concentrated near the origin, 
the criterion is naturally met either at short distances or at small momentum as in Fig.~\ref{fig:Q_Master}. (Bose-Fermi case at the lower temperature column.)

\begin{figure*}[ht]
 \includegraphics[width=12cm]{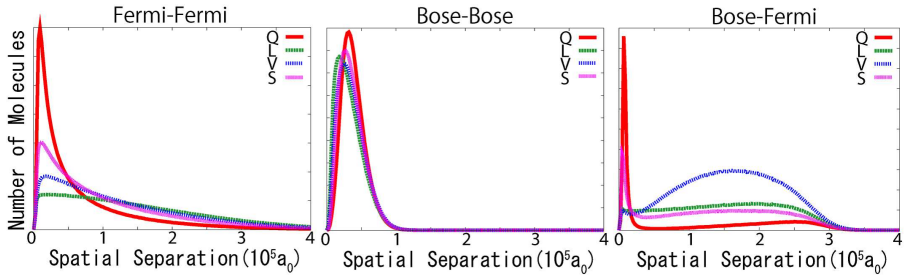}\ \ \ 
  \caption{(Color online) Pair density distributions as functions of spatial separation $\Delta r$ at $T=0.1T_c$. From left to right, Fermi-Fermi, Bose-Bose, and Bose-Fermi. 
  Each solid line pertains to the indicated Criterion. 
  All the peaks representing the initial separation are significantly larger than the experimentally known final size of the molecule.}
  \label{fig:size}
 \end{figure*}

There is another feature worth noting for SPSS.
The likely distance between the atoms in the initial pair turns out to be much larger than the experimentally known final size of the molecule whose diameter is comparable to the scattering length at the final magnetic field. 
To demonstrate this point, Fig.~\ref{fig:size} shows the reduced distribution obtained by summing over the momentum at $T$=0.1$T_c$.
The likely distance is considerably greater for Criteria 2, 3, and 4 than for Criterion 1. 
This observation reflects certain traits of the SPSS approximation. 
The molecular formation takes place within the finite duration of the adiabatic sweep which ChET presumes is sufficiently long for thermalization. 
The SPSS prespecifies the phase-space volume that would be involved in the isothermal transport (which happens to be also ``adiabatic" in SPSS because there is no change in the phase-space volume).
The thermal distribution for each component of the final state can be evaluated at the temperature of the initial state, but then the system cannot be in chemical equilibrium. 
This is where the SPSS and ChET differ, yet the molecular conversion rate comes out surprisingly similar above $T_c$.
Let us note in passing that as the magnetic field is swept, the scattering length grows rapidly near the resonance. 
Across the unitarity limit, the scattering length grows without limit, causing even the widely separated pairs to interact. However, without dynamical calculations, it is not possible to assess the contribution of this resonant region.

Now consider very low temperatures below $T_c$, as we explore the case $T$=0.1$T_c$ in what follows. 
Figs.~\ref{fig:Q_Master}-\ref{fig:S_Master} summarize the phase-space distributions under consideration.
Fig.~\ref{fig:size} plots pair density distributions as functions of the spatial separation $\rho_s(\Delta r)=\int \rho(\Delta r, \Delta p) d\Delta p$.
In the Fermi-Fermi case, the maximum appears near the origin, and then the number of candidate pairs decreases rapidly 
as the spatial and relative momentum separations increase away from the origin.
This trend is independent of the applied criterion, but the height of the peak and the decline in distribution depend on the criterion.
Because of the sparsely spread distribution as seen in Figs.~\ref{fig:L_Master}-\ref{fig:S_Master} away from the origin, 
the initial distance distributions of criteria 2, 3, and 4 are seen to decrease more gradually compared to criterion 1. 
In the Bose-Bose case, pair density distributions $\rho_s(\Delta r)$ show similar tendencies because
two-particle distributions have the same trend, a consequence of the small size of the BEC for all the four criteria.
In the Bose-Fermi case, a structure appears in addition to the peak near the origin. 
In Criterion 1, there are two peaks, one coming from atomic pairs with small spatial separation and 
high relative velocity, and the other one with small relative velocity and large spatial separation 
as noted earlier in relation to Fig.~\ref{fig:ind-1}. 
In the other criteria 2, 3, and 4, they also have double-peaked structure but the profile of these peaks strongly depends on the criterion. 
For example, the outer peak, somewhat broad and round, reaches the highest value for Criterion 3. 
This profile comes from the elliptic-shaped structure seen in the low right panel of Fig.~\ref{fig:V_Master}.
In any combinations of thermodynamic functions, 
all the peaks yield $\Delta r$ much larger than the experimentally estimated final size of the molecule.
SPSS merely enumerates the candidate pairs no subsequent dynamics is implemented.
 
To summarize, we have explored the one- and two-particle distribution functions and have observed the following.
\begin{itemize}
\item The two-particle distribution function behaves more or less the same under all the four criteria examined
except for the Bose-Fermi combination at temperatures below $T_c$.
\item For the Bose-Fermi system, this strong dependence on the imposed criterion is reflected in the molecular conversion
rate.
\item Compared to the molecular size, the peak(s) in the two-particle distribution function corresponds to an
 atomic pair separation which exceeds the known size of the molecule.
\end{itemize}

\subsection{Other Features}
\label{sect:other}

\begin{figure}[h]
\includegraphics[width=8cm]{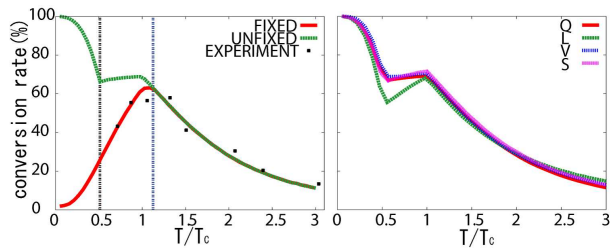}\ \ \ 
  \caption{(Color online) SPSS conversion rates for $^{85}$Rb-$^{87}$Rb using Criteria 1, 2, 3, and 4. 
  The left panel assumes the same temperature for the two components.
  The right one simulates the situation where $^{85}$Rb fails to get cooler at $T$=2.2$T_c$\cite{Papp}.}
  \label{fig:gap-Rb}
\end{figure}

Two aspects of the molecular conversion rate are now analyzed. 
One is the case where there exist two distinct BEC critical temperatures, and the other is the case of a differential gravitational sag that partially separates the two species. We begin with the former. 
As the temperature is lowered, the condensation of the component with higher $T_c$ introduces a decline in the molecular production since the other component still remains in a thermal distribution.
In a real experiment, as was discussed in Ref.~\cite{Papp} with $^{85}$Rb-$^{87}$Rb, the second component fails to attain lower temperature due to technical difficulty and remains thermal. 
In such a case, the decline continues until $T$=0.
See the left panel in Fig.~\ref{fig:gap-Rb}. 
The Fermi-Fermi case is not treated since no conspicuous features arise due to the absence of any phase transition across $T_F$.

\begin{figure}[h]
 \includegraphics[width=8cm]{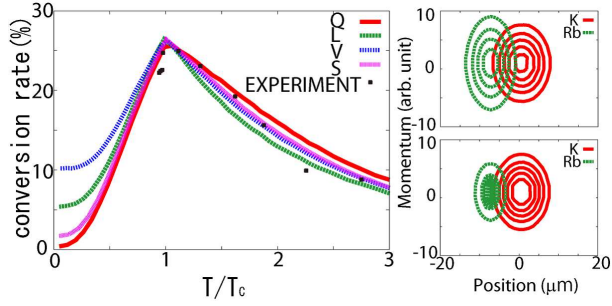}\\
   \caption{(Color online) The left panel shows the SPSS conversion rate for the $^{40}$K-$^{87}$Rb case.
   The right column of panels shows the evolution of the phase-space density distributions for $^{40}$K and $^{87}$Rb as functions of temperature.
   The upper one is for $T>T_c$, and the lower one for $T<T_c$.
The rather large mass difference in $^{40}$K-$^{87}$Rb results in almost nil conversion rate toward $T\rightarrow 0$ for Criterion 1. 
In this figure, the lines of the SPSS molecular conversion rate are scaled by 0.70 to account for molecular losses that occur during rf association; here 
we followed the procedure described in Refs.\cite{olsen,Zirbel}.}
  \label{fig:nil-ovlp}
 \end{figure}

To understand the implications of gravitational sag, it is instructive to see the phase-space density as a function of temperature for $^{40}$K-$^{87}$Rb. 
The BEC phase transition marks the point where the growing spatial gap between the two atomic species leads to 
almost nil conversion rate at lower temperatures. 
See Fig.~\ref{fig:nil-ovlp} for the phase-space density representation of how the overlap evolves as $T$ lowers. 
The gravitational sag can be made insignificant, however, by choosing an appropriate optical trap frequency referred to as the ``magic" frequency in Ref.~\cite{Silke}. 

\subsection{Comparison with ChET}
\label{sect:equ}

The ChET model is now applied to Fermi-Fermi, Bose-Bose, and Bose-Fermi cases with number ratio $\alpha$ set to either 1 or 2/15, and compared with the results of the SPSS model for criterion 1.
The $\alpha=1$ case corresponds to the Fermi-Fermi experiment\cite{Hodby}, and the $\alpha=2/15$ case corresponds to the Bose-Bose experiment\cite{Papp}.

\begin{figure}[h]
 \includegraphics[width=8cm]{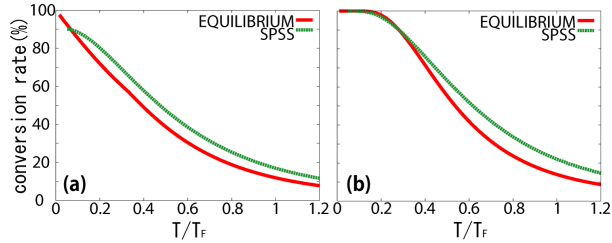}\ \ \
   \caption{(Color online) Temperature dependence of the molecular conversion rate for the Fermi-Fermi case.
   The red line corresponds to the result of ChET, and the green corresponds to the SPSS result.
   (a) the $\alpha$=1 result, (b) the $\alpha$=2/15 result.
   In the figure for $\alpha$=2/15, the characteristic temperature $T_F$ corresponds to the Fermi temperature for the majority atom.}
  \label{fig:equ-ff}
 \end{figure}

The results for the Fermi-Fermi case are shown in Fig.~\ref{fig:equ-ff}.
The ChET produces a conversion rate that increases with decreasing temperature.
And at $T\rightarrow 0$, it always reaches 100\% and is independent of $\alpha$.
SPSS shows a similar trend, but indicates  
a saturation effect at near zero temperature as seen in Fig.~\ref{fig:equ-ff} (a). 
Whether the saturation effect appears or not in the SPSS depends on $\alpha$.

S.~Watabe {\it et. al.} \cite{Shohei2} also applied ChET with 
resonant interaction within the limitations of the mean field approximation. 
Their result indicates that the saturation effect is small. They concluded that the resonant interaction introduces a suppression of
molecule conversion, which reduces the conversion rate somewhat from its maximal rate.

\begin{figure}[h]
 \includegraphics[width=6cm]{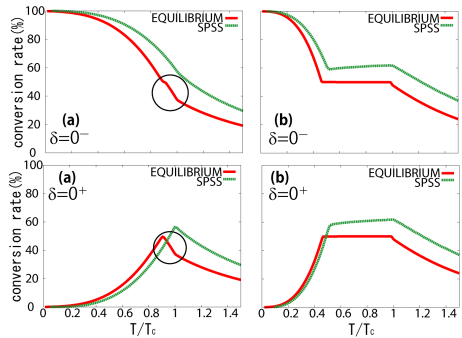}\ \ \
   \caption{(Color online) The same as Fig.~\ref{fig:equ-ff}, except here for the Bose-Bose case.
   Upper panels correspond to $\delta$=0$^-$, lower to $\delta$=0$^+$.
   For the $\alpha$=2/15 case, $T_c$ corresponds to the critical temperature for the major bosonic component.
   In ChET, $T_c$ gets shifted and the plateau regions emerge.}
  \label{fig:equ-bb}
 \end{figure}

\begin{figure}[h]
 \includegraphics[width=6cm]{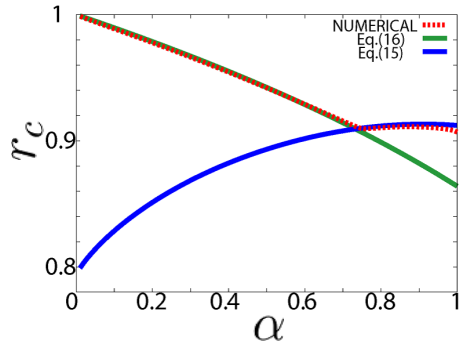}
   \caption{(Color online) Comparison between approximations and a numerical result for the ratio $r_c$.
   At $\alpha\simeq$1, both initial atomic clouds have a BEC component.
   However at $\alpha\simeq$0, only the majority component has a BEC component. 
   This causes differences in the ratio $r_c$ to appear as a function of $\alpha$.}
  \label{fig:gap-bb}
\end{figure}

In the Bose-Bose case, both SPSS and ChET show similar behavior at
thermal temperatures except for a subtle difference stemming from the relationship between the initial temperature $T$ and the
BEC transition temperature $T_c$ in the latter theory.
See the encircled features in Fig.~\ref{fig:equ-bb}. 
ChET assumes that
the system follows the isentropic curve as the magnetic field is swept so that the final temperature
differs from the initial one. 
This causes the difference in the number of atoms in the BEC as reflected in the shift of $T_c$, in contrast to SPSS.
The ratio $r_c$ =$T_c(t_{{\rm final}})$/$T_c(t_{{\rm initial}})$ can be simply evaluated near the two limiting cases.
Setting $T_{0}$ equal to the condensation temperature of the major atomic component at $\delta=$0, and $\mu_{B,0}=\mu_{M,0}$
and applying an approximation to the chemical potential,
we obtain from Eq.\ref{eq:entro}
at $\alpha\simeq $1
\begin{equation}
\frac{8\zeta(4)}{\alpha\zeta(3)}r_c^3 =\frac{2(2-\alpha)\zeta(4)}{\alpha\zeta(3)}+4+\ln \left( \frac{2-\alpha}{\alpha\zeta(3)}\right) - \frac{\alpha\zeta(3)}{8(2-\alpha)},
\end{equation}
and at $\alpha\simeq $0
\begin{equation}
\frac{4\zeta(4)}{\alpha\zeta(3)}r_c^3+\ln \left(\frac{r_c^3}{\alpha\zeta(3)}\right)-\frac{\alpha\zeta(3)}{8r_c^3}=\frac{2(2-\alpha)\zeta(4)}{\alpha\zeta(3)}+\ln \left( \frac{2-\alpha}{\alpha\zeta(3)} \right)- \frac{\alpha\zeta(3)}{8(2-\alpha)}.
\end{equation}
These equations could allow a simple estimation of the shift in $T_c$.
Thus at $\alpha=$1, $r_c$=0.91 and at $\alpha=$2/15, $r_c$=0.99.
Fig.~\ref{fig:gap-bb} compares these approximate analytical results with numerical ones.
This predicted region of a shift in $T_c$, where the trend of the conversion rate suddenly changes, could provide an experimental test for the validity of ChET. 

Moreover, in ChET a plateau appears when the temperature lies
between the two values of $T_c$  
for $\alpha=2/15$ as displayed in Fig.~\ref{fig:equ-bb}.
In ChET, the chemical potential of the majority atoms goes to 0, and
thus the chemical potentials of the rest, {\it i.e.} that of the minority atoms and of molecules must become equal(Eq.~\ref{eq:CPE}).
So in this situation, the conversion rate becomes independent of temperature  
while the rate attained depends on the trap frequency of the minority atoms and that of the molecules.
Below the lower critical temperature, there are two possibilities; one is when 
the atomic BEC is more stable than the molecular BEC, and the other is 
the opposite situation.
In the former case, the SPSS conversion rate decreases with decreasing temperature, 
and it goes to zero at $T\rightarrow 0$, and in the latter case it goes to 100\% as T=0.
And the results of ChET behave similarly.

\begin{figure}[h]
 \includegraphics[width=8cm]{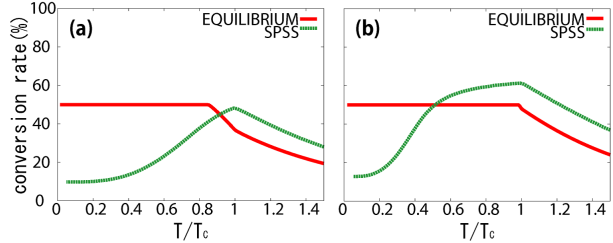}\ \ \
   \caption{(Color online) Temperature dependence of molecular conversion rate for the Bose-Fermi case, the same as in Fig.~\ref{fig:equ-ff}.
   In the ChET, shift of the T$_c$ and the plateau regions appear as in Fig.~\ref{fig:equ-bb}.}
  \label{fig:equ-bf}
 \end{figure}

A marked difference between the two methods appears in the Bose-Fermi case as in Fig.~\ref{fig:equ-bf}.
Consider a system with $\alpha=$2/15 for which there are more bosons than fermions.
The SPSS conversion rate begins to decrease at $T_c$ for both $\alpha=1$ and $2/15$. 
However the rates of descent differ in two cases.
For $\alpha=1$ the decline is steady, but for $\alpha=2/15$,
the SPSS conversion rate is almost flat, decreasing very slowly with decreasing temperature; 
 but it begins to drop suddenly at the temperature 
where the number of fermions exceeds the number of thermal bosons.
But in ChET the conversion rate gets flat-lined below $T_c$.
We emphasize that the temperature dependence of the molecular conversion rate comes out more or less the same for
SPSS and ChET, but a noteworthy difference manifests itself in the Bose-Fermi case below $T_c$, and it would be especially desirable to have more experimental tests available in this regime.

\section{Conclusions}
\label{sect:conclusions}
This work has extended the SPSS analysis of the molecular conversion rate in a quadratic trap, 
including temperature ranges where no experimental explorations have been made to date.
In particular, the temperature dependence of the conversion rate 
below $T_c$ has been considered at great length. This extension has introduced four distinctive
geometrical constraints for the pairing criterion in phase space that controls which atomic pairs contribute significantly to the molecular formation.
Each constraint contains, apart from the geometrical information, a single parameter $\gamma$ which serves to
fix the overall magnitude of the phase-space volume for molecular formation. Our study has examined the sensitivity of the molecular conversion rate
to the geometry and to the magnitude of the parameter $\gamma$. A useful tool for analysis is the quantity we have denoted 
the ``two-particle distribution function", which is the probability distribution of the molecule-forming pairs
in their relative coordinate $|\Delta\vec{r}|=|\vec{r}_A-\vec{r}_B|$ and relative momentum $|\Delta\vec{p}|$ 
defined similarly.
Our investigation shows that once the parameter $\gamma$ is fitted to the experiment at $T$ higher than $T_c$,
then the SPSS method reproduces the experimental results well in the known temperature ranges irrespective of the
constraints (or criteria) used (Sec.\ref{sect:K2}). 

The benchmark system considered is K+K$\rightarrow$ K$_2$ in all the three symmetry combinations Fermi-Fermi, Bose-Bose,
and Bose-Fermi, and the results of the calculations and analysis are the following.
\begin{itemize}
 \item The two-particle distribution function behaves more or less the same under all the four criteria examined
 except for the combination of Bose-Fermi at temperatures below $T_c$.
 \item For the Bose-Fermi system, this strong dependence on the imposed criterion manifests itself in the molecular conversion
 rate.
 \item For all the three cases, the peak(s) in the two-particle distribution function corresponding to the
 separation of a pair of atoms exceeds the known size of the molecule.
 \end{itemize}
Thus, the Bose-Fermi case proved different because the single-particle distribution for the bosonic atoms and that for the fermionic atoms
behave totally differently as they undergo condensation (Sec.\ref{sect:RPS}). Temperatures below $T_c$ remain a challenge for SPSS.
 
The results of SPSS have been compared with those of ChET.
This comparison exhibits a very similar temperature dependence 
of the conversion rate in Fermi-Fermi and Bose-Bose cases.
One marginal difference is that in the Fermi-Fermi case the conversion rate of ChET 
goes to 100\% at $T\rightarrow 0$, 
however in the SPSS model the saturation appears below 100\%.

In the Bose-Bose case, there are two limits as $T\rightarrow 0$, depending on the value of the BEC chemical potential,
{\it i.e.}~exothermic $\delta < 0$ or endothermic $\delta > 0$. The molecular conversion rate tends either to 100\% or to 0\%.
As a result, there is no substantial difference in the temperature dependence of the conversion rate between either theory.
And in the Bose-Fermi case the conversion rate as a function of temperature has a different structure below $T_c$.
ChET yielded a flat-lined conversion rate whereas SPSS gives a slowly temperature-dependent conversion rate
whose behavior is further dictated by the value of $\alpha$.
This paper does not consider effects of inelastic collision which may play an important role below $T_c$.
Further theoretical investigation is needed for deeper understanding of the dynamical behavior of the molecular conversion and an assessment between the theories and the experiments.

This work has revealed what aspects of the atom distributions in phase space affect the conversion
rate and how. The gained information may serve to give an insight when rigorous quantum transport theory 
is effected to visualize the phase-space evolution of the system during the magnetic sweep. After all,
experimental effort for going below $T_c$ to examine the temperature dependence of the conversion rate
appears seriously needed.

This work was supported by Grants-in-Aid for Scientific Research (NO.22340116 and NO.23540465) from the Ministry of Education,
Culture, Sports, Science and Technology of Japan. S.~W.~acknowledges support from a JILA Visiting Fellowship in the summer of
 2010. T.~Y.~acknowledges support from the JSPS Institutional Program for Young Researcher Overseas Visits. The group of C.~H.~G.~ and C.~Z.~ has been supported in part by the U.S. National Science Foundation.


\begin{references}
\bibitem{K2C} In the case of two bosonic isotopes such as $^{39}$K and $^{41}$K used in actual experiments, there are two different condensation temperatures. 
Likewise two fermionic isotopes give rise to two different Fermi temperatures.
Of course, the differences are negligible for our purpose.
\end{references}
\end{document}